\documentclass[reprint,superscriptaddress, amsmath,amssymb,pra, aps,floatfix, longbibliography]{revtex4-1}
\usepackage[utf8]{inputenc}

\usepackage{hyperref}
\usepackage{dcolumn}
\usepackage{bm}
\usepackage{hyperref}
\usepackage[dvipsnames]{xcolor}
\usepackage{graphicx}
\usepackage{siunitx}
\usepackage{lineno}
\usepackage{amssymb}
\usepackage[version=4]{mhchem}
\DeclareUnicodeCharacter{2212}{-}
\graphicspath{ {./images/} }
\usepackage[capitalise]{cleveref}
\usepackage{textcomp}
\usepackage{pdfcomment}
\usepackage{bold-extra}
\usepackage{upgreek}

\DeclareUnicodeCharacter{2009}{\,}
\hypersetup{pdftoolbar=true,        
 pdfmenubar=true,        
 pdffitwindow=false,     
 pdfstartview={fitH},    
 pdftitle={},    
 pdfauthor={Kartik Srinivasan},     
 pdfsubject={vaporPICwaveguides},   
 pdfcreator={Kartik Srinivasan},   
 pdfproducer={LaTeX}, 
 pdfkeywords={Version 1.0},
 colorlinks=true,       
 linkcolor=blue,          
 citecolor=blue,        
 filecolor=blue,      
 urlcolor=blue}

\newcommand{\beginsupplement}{%
 \setcounter{table}{0}
 \renewcommand{\thetable}{S\arabic{table}}%
 \setcounter{figure}{0}
 \renewcommand{\thefigure}{S\arabic{figure}}%
 \renewcommand{\thesubsection}{S-\arabic{subsection}}
}

\usepackage[normalem]{ulem}

\begin{document}

\title{Enabling atom-clad waveguide operation in a microfabricated alkali vapor--photonic integrated circuit}

\author{Rahul Shrestha}   
\thanks{These authors contributed equally}
\affiliation{Joint Quantum Institute, NIST/University of Maryland, College Park, Maryland 20742, USA}
\affiliation{Microsystems and Nanotechnology Division, Physical Measurement Laboratory, National Institute of Standards and Technology, Gaithersburg, Maryland 20899, USA}

\author{Khoi Tuan Hoang}   
\thanks{These authors contributed equally}
\affiliation{Joint Quantum Institute, NIST/University of Maryland, College Park, Maryland 20742, USA}
\affiliation{Microsystems and Nanotechnology Division, Physical Measurement Laboratory, National Institute of Standards and Technology, Gaithersburg, Maryland 20899, USA}

\author{Peter Riley} 
\affiliation{University of Colorado Boulder, Boulder, CO 80303, USA}
\affiliation{Time and Frequency Division, National Institute of Standards and Technology, Boulder, CO 80305, USA}

\author{Roy Zektzer}
\affiliation{Joint Quantum Institute, NIST/University of Maryland, College Park, Maryland 20742, USA}
\affiliation{Microsystems and Nanotechnology Division, Physical Measurement Laboratory, National Institute of Standards and Technology, Gaithersburg, Maryland 20899, USA}

\author{Daron Westly}
\affiliation{Microsystems and Nanotechnology Division, Physical Measurement Laboratory, National Institute of Standards and Technology, Gaithersburg, Maryland 20899, USA}

\author{Paul Lett}
\affiliation{Joint Quantum Institute, NIST/University of Maryland, College Park, Maryland 20742, USA}

\author{Matthew T. Hummon} 
\affiliation{Time and Frequency Division, National Institute of Standards and Technology, Boulder, CO 80305, USA}

\author{Kartik Srinivasan} \email{kartik.srinivasan@nist.gov}
\affiliation{Joint Quantum Institute, NIST/University of Maryland, College Park, Maryland 20742, USA}
\affiliation{Microsystems and Nanotechnology Division, Physical Measurement Laboratory, National Institute of Standards and Technology, Gaithersburg, Maryland 20899, USA}

\date{\today}

\begin{abstract}
    Integrating alkali atomic vapors with nanophotonic devices offers a scalable route to quantum technologies that leverage strong atom-photon interactions. While there have been many approaches to such integration, the general reliance on traditional glass vapor cells, distilled alkali metals, and epoxy sealing limits reproducibility and scalability. Moreover, mitigating adverse Rb-photonics interactions is essential, particularly as devices become more compact and the alkali source lies in close proximity to the photonic elements. Here, we demonstrate the successful operation of compact and fully integrated devices that combine silicon nitride photonic integrated circuits (PICs) with microfabricated borosilicate vapor cells and pill-type rubidium (Rb) dispensers through hermetic seals via anodic bonding. We show how successful operation hinges on optically activating the dispenser in a low-power pulsed mode, releasing controlled amounts of Rb vapor on demand while mitigating photonic degradation. Simultaneously, a counter-propagating desorption laser completely suppresses Rb-induced losses and enables waveguide-based atomic vapor spectroscopy. Using this approach, we demonstrate repeatable control of vapor density by tuning activation pulse length, duty cycle, and device temperature. These results establish a compact, manufacturable, and scalable vapor--PIC device, and set the stage for future demonstrations in cavity quantum electrodynamics, quantum nonlinear optics, and chip-scale atomic sensors.

\end{abstract}

\maketitle

\section{Introduction}
Integrating thermal atomic gases with photonic circuits paves the way for compact, chip-scale devices incorporating light sources, atomic media, routing, filtering, interference, and detection, thereby enabling deployable atom–light–interaction-based technologies such as precision sensors, frequency references, and nonlinear optical systems. Alkali atoms such as rubidium (Rb) are of particular interest due to the wide range of experiments and applications they support~\cite{kitching_chip-scale_2018}. Various systems have been put forth to enable interactions of alkali vapors with guided optical waves, including tapered nanofibers~\cite{spillane_observation_2008,salit_ultra-low_2011,jones_saturation_2014,jones_ladder-type_2015,song_absorption_2019,lamsal_transmission_2019,finkelstein_super-extended_2021}, hollow-core fibers~\cite{ghosh_low-light-level_2006,londero_ultralow-power_2009,slepkov_spectroscopy_2010,venkataraman_few-photon_2011,sprague_efficient_2013,perrella_high-resolution_2013,donvalkar_frequency_2014,perrella_engineering_2018}, hollow-core waveguides \cite{schmidt_atomic_2010}, and photonic integrated circuit (PIC) waveguides~\cite{yang_atomic_2007,wu_slow_2010,stern_nanoscale_2013,ritter_atomic_2015,stern_strong_2017,ritter_coupling_2018,zektzer_nanoscale_2021,zektzer_strong_2024}. PIC-based devices have recently attracted particular interest due to the varied ways in which they control optical fields~\cite{yang_atomic_2007,stern_nanoscale_2013,ritter_coupling_2018,davidson-marquis_coherent_2021}, the ability to realize strongly enhanced light-matter interactions in PIC resonators and waveguides~\cite{stern_strong_2017,ritter_coupling_2018,zektzer_strong_2024}, the straightforward collection of light that has interacted with the atoms, and the potential for integration with complementary components, particularly in silicon nitride photonics~\cite{poon_silicon_2024,perez_high-performance_2023,moille_integrated_2022,zhou_prospects_2023,isichenko_multi-laser_2025}. Ideally, hybrid vapor--PIC systems should be created through scalable fabrication processes, using an alkali metal source that is straightforward to handle and incorporate within such a fabrication flow. 


In ref.~\cite{riley_evanescent_2025}, we demonstrated a process for creating fully-integrated vapor--PIC devices, where silicon nitride PICs are anodically bonded to borosilicate glass vapor cells that incorporate Rb `pill' dispenser sources (a similar effort has also recently been reported~\cite{grosman_wafer-scale_2025}). Such Rb sources, which consist of a mixture of  rubidium molybdate and zirconium and aluminum powders, are commercially available, atmosphere stable, and require only a simple infrared laser activation mechanism for releasing Rb after completed fabrication. As a result, they are now commonly used in applications such as atomic clocks, magnetometers, and wavelength references~\cite{loh_microresonator_2016,sebbag_chip-scale_2021,knapkiewicz_dynamically_2019,han_microfabricated_2018,hummon_photonic_2018,stern_chip-scale_2019,lucivero_laser-written_2022,januszewicz_chip-scale_2025}. 

Using the above platform, in this work, we focus on near-field interactions between Rb vapor and the evanescent field of air-clad PIC waveguides. We demonstrate the operation of such vapor--PIC waveguides by employing multiple strategies to limit the impact of adverse Rb-PIC interactions. We start by considering the most straightforward approach, in which the Rb pill source is activated with a standard process similar to that used in the above applications, and show that it results in a significant increase in the propagation loss of the air-clad PIC waveguide ($>\SI{1750}{\decibel\per\centi\meter}$), so that only very short waveguides (on the order of $\approx$ \SI{100}{\micro\meter} in length) are viable for waveguide-based Rb spectroscopy. We then consider methods to overcome such Rb-induced PIC loss, through experiments that study simultaneous Rb pill activation, free-space Rb spectroscopy, and PIC waveguide spectroscopy. We demonstrate that the combination of low-power pill activation with a counter-propagating desorption laser sent through the waveguide results in waveguide losses that are not degraded by pill activation (to within our measurement uncertainty), thereby enabling waveguide-coupled Rb spectroscopy. Moreover, we characterize the attributes of this low-power activation approach, which consists of a transient decay of the spectroscopic signal that depends on laser pulse power and duration. We show that the Rb vapor density can be controlled by laser activation parameters, with cell temperature providing additional control, and that through repetitive pulses, a quasi-steady density regime can be achieved. Our results are of particular relevance to quantum optics within vapor--PIC systems, including recent studies pointing to near-term opportunities in strong coupling cavity quantum electrodynamics between single photon and single vapor-phase atoms~\cite{alaeian_cavity_2020,zektzer_strong_2024,austin_vapor-cavity-qed_2025,larsen_chip-scale_2025}.

\begin{figure}[ht]
\centering
\includegraphics[width=\linewidth]{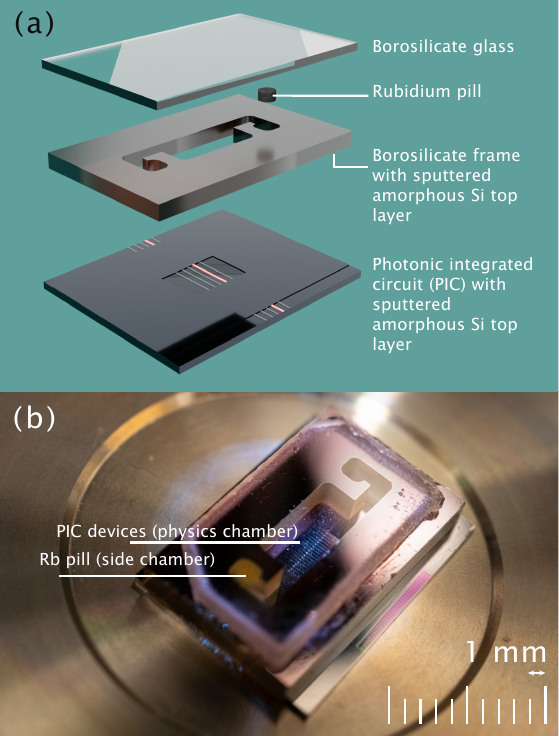}
\caption{\textbf{Hybrid device that integrates an atomic vapor cell, Rb pill source, and a photonic integrated circuit.}
\textbf{(a)} A borosilicate cover glass, micro-machined borosilicate frame, and a photonic integrated circuit (PIC) are anodically bonded together along with a miniature Rb pill to form a compact “vapor-PIC” package. \textbf{(b)} A photograph of a device after bonding and Rb pill activation.}
\label{fig:conceptDevice}
\end{figure}

\section{vapor--PIC device}
The devices are fabricated according to the process described in detail in Riley et al. \cite{riley_manuscript_2025}. Briefly, each device consists of three main units that are assembled via anodic bonding and shown schematically in Fig.~\ref{fig:conceptDevice}(a). From bottom to top, we have the PIC, which serves as the base of the device, an ultrasonic machined borosilicate glass frame, which forms the cell walls, and a borosilicate glass lid, which forms the cell top. The layers are sequentially anodically bonded to form a hermetically sealed vacuum cell device with dimensions of $\SI{12}{mm} \times \SI{16}{mm} \times \SI{3.5}{mm}.$ Fig.~\ref{fig:conceptDevice}(b) shows a photograph of a completed device. During this process, a \ce{Rb2MoO4} Zr/Al pill-type dispenser for generating Rb vapor (upon laser activation) is placed in a side chamber which is connected to the center chamber with a right angle channel. The dispensed Rb vapor contains both naturally occurring $^{85}\text{Rb}$ and $^{87}\text{Rb}$ isotopes.  

\begin{figure*}
\centering
\includegraphics[width=\textwidth]{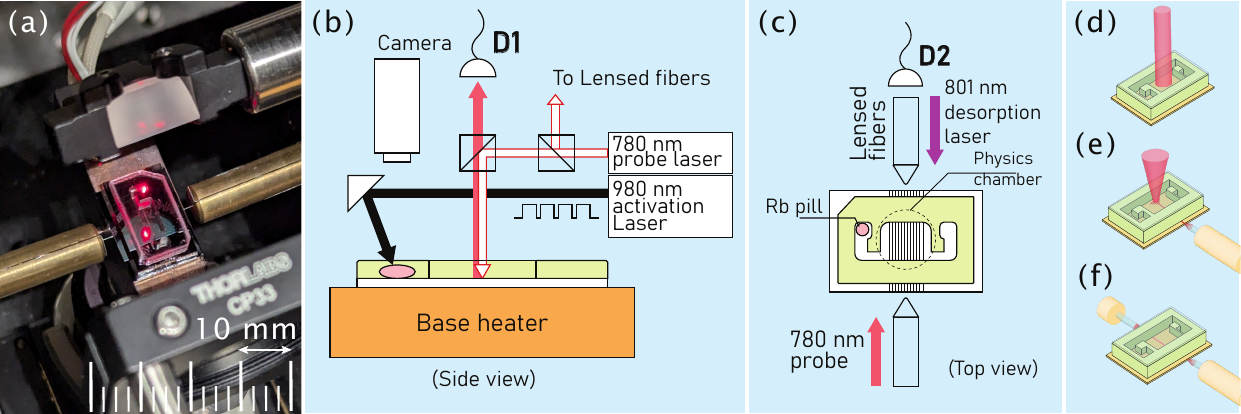}
\caption{\textbf{Experimental setup and schematic for simultaneous Rb pill activation, Rb density assessment, and PIC waveguide absorption spectroscopy.}
\textbf{(a)} \textit{Photograph of a device in setup}: The device rests on a heated copper sample mount while the lensed fibers couple light in and out of the device. A D-mirror and lens guide the 980 nm activation beam onto the Rb pill.
\textbf{(b)} \textit{Side view schematic}: the device (green/white) containing the Rb pill (pink) rests on top of a temperature-controlled copper base heater (orange). A \SI{980}{nm} activation laser (black arrow) is amplitude-modulated, as indicated by the square-wave inset, and focused onto the pill under a microscope camera. A \SI{780}{nm} probe laser is split (50:50) with one arm  coupled through a lensed fiber into the PIC waveguide, and the other interrogating the free-space Rb vapor. Probe light reflected from the chip exits upward to detector D1.
\textbf{(c)} \textit{Top view schematic}: two lensed fibers couple light in and out of the PIC waveguide (green/white): the \SI{780}{nm} probe (red arrow) and a counter-propagating \SI{801}{nm} desorption laser (magenta arrow). Probe transmission is measured by detector D2, while the desorption beam prevents Rb buildup on the waveguide during pill activation. The region of the device where the waveguides are air-cladded and interact with the Rb vapor is highlighted with dashed circle. 
\textbf{(d-f)} \textit{Illustrations showing three different measurement methods}: free-space interrogation of Rb using reflection off the chip (path length of \SI{4}{mm}), free-space interrogation using a fully oxide-clad, waveguide-coupled grating out-coupler (path length of \SI{2}{mm}), and interrogation by the waveguide evanescent field (interaction length of \SI{60}{\micro\meter} or \SI{3}{mm}).}
\label{fig:setup}
\end{figure*}

The PIC waveguides are created in a \SI{250}{nm} thick silicon nitride layer created through low pressure chemical vapor deposition, electron-beam lithography, and reactive ion etching. A selective \ce{SiO2} top cladding surrounds the entire PIC layer apart from regions where evanescent interactions with the Rb vapor are desired~\cite{zektzer_strong_2024}. We use a chemical mechanical polishing process to smoothen and planarize the oxide-clad regions, which are additionally coated with amorphous silicon, so that they can support a hermetic seal during subsequent anodic bonding. Devices contained on the $\SI{12}{\milli\meter} \times \SI{16}{\milli\meter}$ die include 600~nm width waveguides with \SI{60}{\micro\meter} and 3~mm long air-clad regions and fully oxide-clad waveguide-coupled grating out-couplers. The gratings have a uniform period of \SI{440}{nm} and are simulated to diffract light off the chip at an angle of \SI{10}{\degree} with respect to normal incidence. These photonic devices interact with Rb vapor in the center chamber designated as the physics chamber in Fig.~\ref{fig:conceptDevice}(b). 

Before discussing the operation of the above devices in detail, we provide some context with respect to other vapor--PIC devices that have been studied in recent years. In general, one can describe such devices based on a number of key characteristics, including: (1) PIC device designs, (2) vapor--PIC integration approach (i.e., bonding technology) and size, and (3) alkali vapor source. Both early work on chip-integrated hollow-core waveguides~\cite{yang_atomic_2007,wu_slow_2010} and more recent studies with PIC SiN waveguides~\cite{stern_evanescent_2013,stern_enhanced_2016,zektzer_strong_2024} have utilized epoxy seals to hand-blown gas cells that are filled via Rb distillation. Anodic bonding approaches have also been used, both in a triple-stack configuration in which the cell frame is fused to a glass tube leading to a Rb reservoir~\cite{ritter_coupling_2018,skljarow_purcell-enhanced_2022}, as well as in configurations in which the PIC is bonded to a larger glass cuvette that is attached to a UHV apparatus and alkali source for cold atom applications~\cite{mcbride_demonstration_2025}. In contrast, we combine PIC waveguides with a standard anodic bonding process that enables incorporation of a compact vapor cell and commercially available Rb pill source. Our standalone devices have an approximate volume of \SI{57}{\milli\meter\cubed} that is more than one order of magnitude smaller than the $\approx \SI{1480}{\milli\meter\cubed}$ volume of epoxy-bonded devices incorporating small glass-blown cells~\cite{zektzer_strong_2024}. We note that similar compact PIC-vapor cell devices made through anodic bonding have recently been reported~\cite{grosman_wafer-scale_2025}; however, as discussed in the next section, both works suffer from waveguide degradation under typical Rb pill activation. 

\section{Challenges with rubidium pill - PIC integration \label{section:60micrometer}}

\begin{figure}
    \centering
    \includegraphics[width=\linewidth]{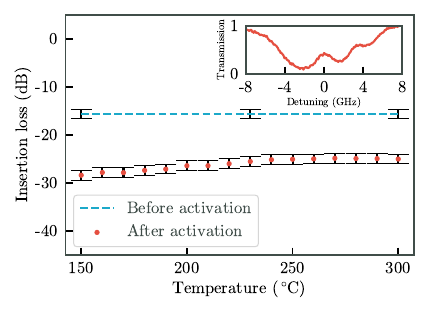}
    \caption{\textbf{Loss induced by standard activation that results in a stable vapor density.} Temperature dependence of the total insertion loss at \SI{780}{nm}, away from any Rb transition, of a \SI{60}{\micro\meter} long PIC waveguide before and after Rb is introduced. The Rb vapor is generated using a Rb pill dispenser after heating with a strong \SI{980}{nm} external laser ($\approx \SI{1}{W}$). After Rb introduction, there is an $\approx \SI{10.5}{\decibel}$ increase in insertion loss of the waveguide. \textbf{Inset:} Doppler-broadened absorption spectrum of the $D_2$ manifold through the waveguide at \SI{240}{\celsius}.
    The error bars represent \SI{1}{\decibel} measurement uncertainty, and are one standard deviation values associated with coupling variation.}
    \label{fig:steadystate}
\end{figure}

We begin by considering the operation of the air-clad PIC waveguides when the Rb dispenser pill is laser activated according to a standard procedure utilized for vapor cell clocks, magnetometers, and related technologies~\cite{douahi_vapour_2007,giridhar_mems_2022,jia_microfabricated_2022}. In this procedure, hereafter referred to as standard activation, the Rb pill is activated using a high power ($\approx\SI{1}{W}$) heating laser at \SI{980}{nm}, focused on the pill for 5~s to 10~s, to dispense Rb vapor into the cell. It is then inserted in an experimental setup in which a 780~nm tunable laser is coupled into and out of the device with lensed optical fibers (similar to that shown in Fig.~\ref{fig:setup}(a)), and we start by considering device transmission when the laser is tuned off-resonance of any Rb transition (confirmed using a reference Rb cell). Unfortunately, we find that standard activation of the Rb pill results in near-total loss of transmission for the waveguides with 3~mm long air-clad region. This is presumably caused by the activation having released a significant amount of Rb that sticks onto the PIC waveguides and increases their propagation loss, in particular considering the short ($\lesssim$ 5~mm) separation between the pill and waveguide regions. However, short waveguides with a length close to \SI{100}{\micro\metre}, exhibit non-zero transmission. Their insertion loss increases significantly, by a magnitude of $\approx \SI{10.5}{\decibel}$ for a \SI{60}{\micro\metre} waveguide (\SI{1750}{\decibel\per\centi\meter}). Here we define the insertion loss as $10 \times \log(P_{out}/P_{in})$ where $P_{in}$ and $P_{out}$ are power through the input facet and power out of the output facet of the waveguide, respectively. We plot the insertion loss of the waveguide as we adjust the device temperature from \SIrange{150}{300}{\celsius} in Fig.~\ref{fig:steadystate}. Though the loss remains roughly constant for different temperatures after Rb pill activation, we do observe a weak correlation between the higher temperature and lower losses. This suggests that after the first drastic increase in loss, there is no further increase in loss with more Rb in the vapor phase, and rather, that higher temperatures may enable some desorption of the Rb. 

An inset in the same plot also shows the Doppler-broadened absorption spectrum of the $D_2$ manifold at \SI{240}{\celsius} taken through the \SI{60}{\micro\meter} long waveguide. Despite the high temperature, the spectroscopy laser is not completely absorbed at the atomic resonances due to the short interaction length, though the high temperature also results in a large Doppler broadening of the lines. A stronger contrast at lower temperatures requires longer interaction lengths and hence operation of the waveguides with longer air-clad regions. Given the losses observed above, realizing operation of the longer waveguides within the current device architecture requires a more careful evaluation of the Rb pill activation process. A similar conclusion was made in ref.~\cite{grosman_wafer-scale_2025}, where large PIC waveguide propagation losses were suffered after stable Rb pill activation, limiting device lengths to a few hundred micrometers at most.

\section{Controlled rubidium pill activation}
\label{section:controlled}
The above results suggest that limiting the amount of Rb released from the pill during activation may limit PIC degradation. Of course, it is also essential that the activation release sufficient Rb for spectroscopy experiments. To balance these considerations, in this section we first describe an experimental apparatus that enables Rb pill activation while simultaneously monitoring the free-space Rb absorption signature and the PIC device performance. We then describe how low-power activation of the pill allows for operation in a transient regime, how repetitive pulsed operation of the activation laser maintains a quasi-steady Rb density, and how the cell temperature affects Rb density.

\subsection{Setup for PIC monitoring during Rb pill activation \label{section:setup}}

The fabricated device is interrogated using the setup whose schematic is shown in Fig.~\ref{fig:setup}. The sample sits on a temperature-controlled copper base heated above $\SI{100}{\celsius}$ in an attempt to prevent Rb from sticking to the PIC. A \SI{980}{nm} "activation laser" is focused onto the Rb pill using a $f=\SI{50}{mm}$ lens to a 
spot diameter of $\approx \SI{1}{mm}$. By adjusting the angle of a D-mirror, as confirmed by an imaging system fixed above the device, we direct the beam on the pill (Fig.~\ref{fig:setup}(a)-(b)). We employ a two-stage activation process where $\approx \SI{1}{\watt}$ of power is used to initially release the first burst of Rb vapor, with subsequent activation steps using $\approx \SI{100}{mW}$ of power to controllably release Rb. The activation laser is modulated at this power for pulsed release of Rb vapor, as we will describe further below. A \SI{780}{nm} probe laser is split for free-space and waveguide spectroscopy of the Rb vapor. Two lensed fibers are used to couple light in and out of the waveguide while the other half of the light is reflected off the chip surface for free-space spectroscopy. In addition to the free-space spectroscopy from above the chip (Fig.~\ref{fig:setup}(d)), we also have waveguide-coupled on-chip gratings that can project light upwards, through the cell (Fig.~\ref{fig:setup}(e)). We also employed a \SI{801}{nm} desorption laser through the waveguide, counter-propagating relative to the \SI{780}{nm} probe (Fig.~\ref{fig:setup}(c)), to prevent Rb adhesion to the PIC waveguides (to be discussed in section \ref{section:desorption}).

\subsection{Low-power activation and transient Rb density \label{section:transient}}

\begin{figure*}[ht]
\centering
\includegraphics[width=\textwidth]{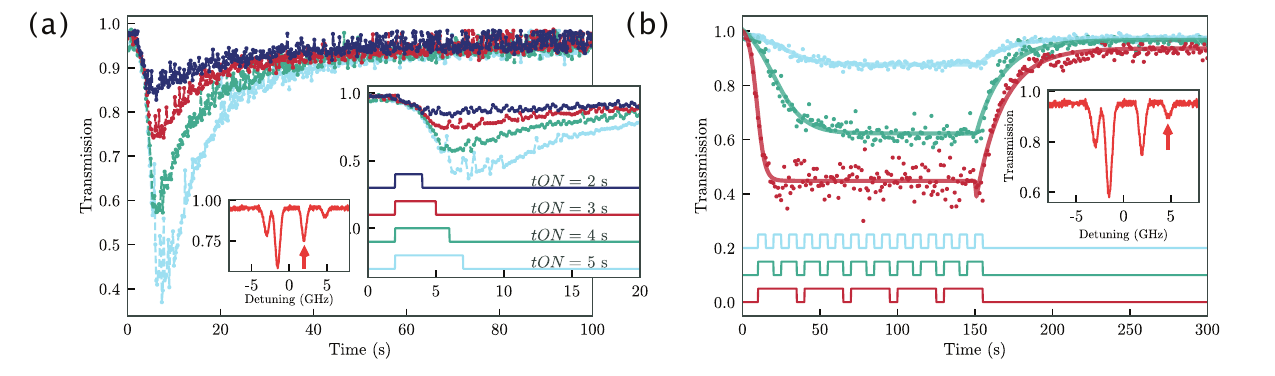}
\caption{\textbf{Rubidium density control via activation laser.}
\textbf{(a)} \textit{Transient Rb density.} Transmission of the $5S_{1/2},F=2 \rightarrow 5P_{3/2} $ transition in \ce{^{85}Rb} is monitored for 100~s in free-space (configuration in Fig.~\ref{fig:setup}(d)) while a \SI{980}{nm} activation laser dispenses Rb from a pill. Four activation lengths -- \SI{2}{s} (dark blue), \SI{3}{s} (red), \SI{4}{s} (green), and \SI{5}{s} (light blue) -- are compared at identical laser power ($\approx \SI{100}{mW}$) and device temperature of \SI{180}{\celsius}. Data points (symbols) are joined by dashed lines to guide the eye. Longer activation pulse lengths ($tON$) yield higher Rb vapor densities, seen as deeper transmission dips: the 2 s trace (dark blue) maintains the highest transmission, whereas the 5 s trace (light blue) shows the lowest.
\textbf{Upper right inset}: a 0~s to 20~s zoom-in of the data reveals an exponential drop in transmission (linear rise in Rb density) during each ON window and an exponential recovery when the laser is OFF; solid square pulses beneath the traces indicate the programmed 2 s, 3 s, 4 s, and 5 s activation windows.
\textbf{Lower left inset}: full Doppler-broadened absorption spectrum of the $D_2$ manifold from a reference cell; the arrow marks the hyperfine line whose transmission is tracked in the main panel.
\textbf{(b)} \textit{Quasi-steady Rb vapor density via repetitive operation.}
Transmission of the $5S_{1/2},F=1 \rightarrow 5P_{3/2} $ transition in \ce{^{87}Rb} is monitored for \SI{300}{s} in free-space while a \SI{980}{nm} activation laser periodically releases Rb from a pill. Three programmed duty cycles --- low (light blue), medium (green) and high (red) --- are shown as solid square-wave traces beneath the data (high level = laser ON). During the pulsing window (0~s to 150~s), a nearly constant Rb density is achieved. When pulsing stops at $t\approx\SI{150}{s}$, the vapor gradually re-adsorbs onto the cell walls and all traces return toward unit transmission. The transmission during the pulsing period is fitted with a four-parameter logistic function and the transmission after pulsing stops is characterized by an exponential. Further discussion about the fitting is present in the main text. \textbf{Upper right inset}: Doppler-broadened absorption spectrum of the $D_2$ manifold; the arrow marks the hyperfine line tracked in the main panel.}
\label{fig:RbCtrl}
\end{figure*}

As discussed earlier, standard activation of a Rb dispenser pill usually involves application of a single high power laser pulse that deposits a Rb reservoir in an auxiliary chamber connected to the physics chamber. The initial high power pulse also generates enough Rb to coat the inner glass walls with several monolayers of Rb~\cite{ma_modification_2009},  resulting in the cell's equilibrium Rb vapor pressure being governed by the cell temperature. Unfortunately, this single high-power activation severely degrades our photonic devices, as evidenced by a catastrophic increase in insertion loss shown in section \ref{section:60micrometer}.

To avoid that damage, we adopt a two-stage protocol. In stage 1, we apply \SI{1}{\second} laser pulses with increasing power from \SI{100}{mW} to $\approx \SI{1}{W}$ while simultaneously monitoring the free-space Rb absorption spectrum from above (see Section \ref{section:setup} and Fig. \ref{fig:setup}). There is an initial burst of Rb released from the pill, which is detected through spectroscopy. This initial release is detected closer to \SI{1}{W} of optical power depending on the specific focusing conditions based on pill orientation and placement. This careful activation based on monitoring the Rb spectroscopy signature while gradually increasing the laser power and working with a pulse duration that is much shorter than that used in the standard activation in Section~\ref{section:60micrometer} allows us to obtain the minimum amount of Rb in our cell. The Rb vapor rapidly adsorbs onto the cell walls. In stage 2, we reduce the power of the activation laser to $\approx \SIrange{100}{200}{\milli\watt}$. Within this power range, we are able to release an arbitrary amount of Rb into the cell by changing the length of the activation pulse.  As the pulse ends the Rb vapor density in the cell decays to zero.

To study the transient behavior in more detail, we followed the temporal behavior of the transmission of the  $5S_{1/2},F=2 \rightarrow 5P_{3/2} $ transition in \ce{^{85}Rb}, measured in free-space, across an interval of 100~s and with a temporal resolution of \SI{0.2}{s} as shown in Fig.~\ref{fig:RbCtrl}(a).  With laser power of $\approx \SI{100}{mW}$ and device temperature at \SI{180}{\celsius}, we took four different sequences at varying activation pulse lengths of \SI{2}{s}, \SI{3}{s}, \SI{4}{s}, and \SI{5}{s}. Unsurprisingly, longer pulse time yield higher densities of Rb, as reflected by a larger maximum transmission contrast (lower transmission value). The transmission exponentially decreases during the duration of the laser pulse and subsequently there is an exponential restoration with $1/e$ time of \SI{\approx 10}{\second} in transmission after the pulse ends. 

\subsection{Achieving quasi-steady Rb density with repetitive activation \label{section:pulsed}}
Having observed how the Rb density behaves in response to a single activation pulse of varying duration, we next study its behavior in response to a series of pulses. Each activation pulse at the relatively low-power of $\approx \SI{100}{mW}$ releases a small amount of Rb which rapidly decays. We modulate the amount of time the laser is on ($tON$) and off ($tOFF$) to achieve a quasi-steady density of Rb. 

In Fig.~\ref{fig:RbCtrl}(b), we probe the transmission of the $5S_{1/2},F=1 \rightarrow 5P_{3/2} $ transition in \ce{^{87}Rb} for 300~s with the free-space configuration. We show three activation laser duty cycles with $tON=$ 1~s, 4~s, and 9~s and $tOFF=$ 1s, and the total pulse sequence length of 150~s. The laser power and temperature of the device are kept constant. During the pulsing window, every $tON$ raises the vapor density (reducing transmission), whereas each $tOFF$ lets the density decay (restoring transmission). Larger $tON$ times result in accumulation of larger Rb density, leading to a higher quasi-steady density of Rb.

The transmission during the pulsing window (0~s to 150~s) is fit with a four-parameter logistic function 

\begin{equation}
\alpha(t)= \alpha_0 + \frac{\alpha_{f}}{1+e^{-k(t-t_0)}},
\label{eq:alpha(t)}
\end{equation}

\noindent where $\alpha_0$ is the initial transmission, $k(T)$ is temperature-dependent reaction rate, $\alpha_f$ is the stable transmission when Rb density stabilizes, and $t_0$ is when the pulses start. Using the Beer-Lambert law, we can extend it to express the density of Rb atoms over time, 

\begin{equation}
N(t)=-\frac{1}{\sigma L}\ln\left(\alpha_0 + \frac{\alpha_{f}}{1+e^{-k(t-t_0)}}\right),
\label{eq:N(t)}
\end{equation}

\noindent where $\sigma$ is the atomic cross section and $L$ is the interaction length. Since logistic reaction models or Prout-Tompkins (PT) models have been used to model chemical reactions involving thermal decomposition \cite{burnham_use_2017}, it is appropriate for characterizing the decomposition of the Rb pill by the activation laser. Both the temperature of the pill, which is determined by experimental conditions like the duty cycle and the power of the activation laser, and overall cell temperature impact the Rb density, as we explore in the next section.  

Extensive chemical kinetics study of the Rb pill with energy provided by the laser pulses is possible but beyond the scope of this paper. However, our rudimentary analysis using the PT model, equation \ref{eq:alpha(t)}, gives us an estimate of the interval over which we can expect a quasi-steady density. This is of greater importance for higher densities of Rb where the variance in transmission within the nominally flat transmission region is considerably higher. For instance, the three duty cycles shown in Fig.~\ref{fig:RbCtrl}(b) have mean transmission of 0.88(1), 0.62(3), and 0.45(5).

\subsection{Influence of cell temperature \label{section:temp}}
\begin{figure}[ht]
\centering
\includegraphics[width=\linewidth]{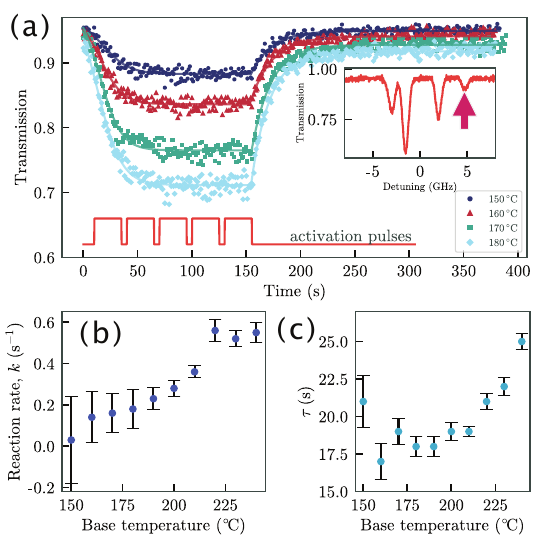}
\caption{\textbf{Temperature-dependent Rb loading and unloading during low power activation.}
\textbf{(a)} Transmission of the $5S_{1/2},F=1 \rightarrow 5P_{3/2} $ transition in \ce{^{87}Rb} is monitored for \SI{400}{s} in free-space using an on-chip grating coupler (Fig.~\ref{fig:setup}(e)) while a \SI{980}{nm} activation laser periodically releases Rb from a dispenser pill. The solid square-wave trace beneath the data marks the laser timing (high = ON). The device base is held at \SI{150}{\celsius} (dark blue circles), \SI{160}{\celsius} (red triangles), \SI{170}{\celsius} (green squares), and \SI{180}{\celsius} (light blue diamonds); laser power and duty cycle remain fixed. Scattered points are data; solid curves are fits to the logistic model for the time interval between \SI{0}{s} and \SI{150}{s}) followed by a single-exponential recovery for all times $>\SI{150}{s}$. \textbf{Inset:} Doppler-broadened $D_2$ absorption spectrum recorded in a reference cell; the arrow indicates the hyperfine transition tracked in panel a.
\textbf{(b)} Reaction rate $k$ extracted from the \SI{0}{s} to \SI{150}{s} fits versus base temperature. 
\textbf{(c)} Transmission-decay rate constant obtained from fitting the $>\SI{150}{s}$ data vs. base temperature.
Error bars in panels (b) and (c) denote one standard deviation uncertainties from the fits.
}
\label{fig:tempDepPulsed}
\end{figure}

In section \ref{section:pulsed} we showed that control of the quasi-steady density of Rb can be achieved by pulsing the Rb pill with differing $tON$ times at fixed (low) activation laser power and device temperature. Now we study the influence of cell temperature on this quasi-steady density and compare against the standard Clausius-Clapeyron (CC) model \cite{gianni_di_domenico_vapor_2011} used to characterize atomic vapors. 

We monitor the transmission of the $5S_{1/2},F=1 \rightarrow 5P_{3/2} $ transition in \ce{^{87}Rb} using free-space spectroscopy with the help of on-chip gratings while we pulse the activation laser to achieve a quasi-steady density of Rb (Fig.~\ref{fig:setup}(e)). However, instead of varying the laser on time ($tON$), we vary the temperature of the device while keeping the laser power ($\approx \SI{100}{mW}$) constant. Fig.~\ref{fig:tempDepPulsed}(a) shows the transmission for a duration of \SI{400}{s} at four different temperatures: \SI{150}{\celsius}, \SI{160}{\celsius}, \SI{170}{\celsius}, and \SI{180}{\celsius}. During the pulsing window (\SI{0}{s} to \SI{150}{s}), the Rb density rises and the transmission dips following the logistic model discussed in Section~\ref{section:pulsed} and reaches a stable plateau. However, increasing the device temperature increases the reaction rate, which allows the Rb to reach a higher density at higher temperatures. In Fig.~\ref{fig:tempDepPulsed}(b), we plot the reaction rates extracted from the logistic model fitted to the transmission during the pulsing period. The trend is roughly exponential from \SI{150}{\celsius} to \SI{220}{\celsius}, following the typical Arrhenius equation where $\ln(k(T))\propto -1/T$; however, after \SI{220}{\celsius}, we see signs of plateauing. After the pulsing stops (>150 s), the transmission recovers. The recovery is no longer governed by equation \ref{eq:alpha(t)}; instead, we fit an exponential function, $e^{-t/\tau}$, and extract the time constants $\tau$, where $\tau$ can be understood as a characteristic timescale for the speed of Rb vapor pumping by the glass walls. We plot $\tau$ as a function of temperature in Fig.~\ref{fig:tempDepPulsed}(c). We see a rough trend where $\tau$ increases after \SI{200}{\celsius}, which implies the pumping efficiency is slightly lowered at higher temperatures.  

\begin{figure}[h!]
\centering
\includegraphics[width=\linewidth]{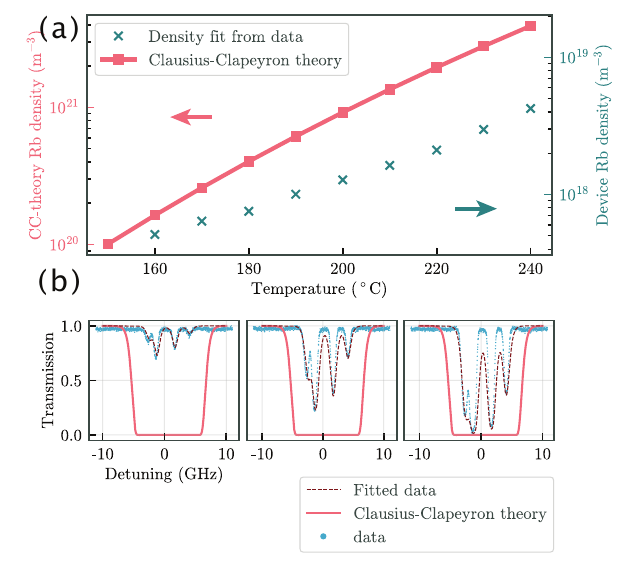}
\caption{\textbf{Temperature‑dependent Rb density: comparison with a solid-vapor equilibrium model.}
(a) The quasi‑steady vapor density obtained in the low power pulsed protocol of Fig.~\ref{fig:RbCtrl}(b) is measured by free‑space absorption spectroscopy above the chip (Fig.~\ref{fig:setup}(d) configuration). The relative variation of the experimental density (cyan) is plotted against temperature and compared with the standard Clausius–Clapeyron (CC) scaling (red). Deviations grow with temperature, indicating that this operation of the pill source departs from equilibrium‑vapor behavior. The left and right scale show the absolute Rb density for experiment and theory, respectively. Experimental densities are more than two orders of magnitude lower than CC values. (b) Representative spectra at \SI{160}{\celsius}, \SI{200}{\celsius}, and \SI{240}{\celsius} are shown from left to right. The cyan data points are the measurements, dashed lines are least‑squares fits to the data using a CC model, and the solid red traces are the CC model predictions under standard vapor equilibrium conditions. The experimental data show dips that are noticeably narrower than the fits based on the CC model.}
\label{fig:compCC}
\end{figure}

Generally, the vapor pressure or the density of Rb vapor in a thermodynamic equilibrium with its solid, for instance in a conventional vapor cell, is purely a function of temperature. The vapor pressure is given by the CC relation where the Rb density is dependent on the temperature of the cell. It is clear that we deviate from this behavior as we have shown activation laser power and time to be effective knobs to control Rb density at a constant temperature. Temperature still plays a role to control the absolute Rb density; but ultimately, as shown in Fig.~\ref{fig:compCC}(a), we can be orders of magnitude below the expected equilibrium Rb density at a particular temperature. In the same figure, we show that the relative increase in density with temperature (at a starting point of \SI{150}{\celsius}) also diverges from the standard CC model. In Fig.~\ref{fig:compCC}(b), we display representative spectra at \SI{160}{\celsius}, \SI{200}{\celsius}, and \SI{240}{\celsius}, where the cyan dots show experimental data for our low power, pulsed Rb dispenser source operation, the red dashed lines show the best fits of this data to a CC model, and the red solid traces are the Doppler broadened spectra expected at those temperatures under standard equilibrium conditions with the CC model. The experimental data indicates that within our low power activation regime, there is significantly lower Rb density than the prediction from the standard CC model under vapor equilibrium. That being said, when fit to a CC model, the transmission dips in the experimental data can still be well-aligned with the dips of the fits; however, they are narrower. This suggests that the velocity distribution of the Rb atoms released from the pill also deviates from the Rb vapor at equilibrium.    

\section{PIC Waveguide Operation with the Pulsed rubidium Source \label{section:desorption}}
\begin{figure}[ht]
    \centering
    \includegraphics[width=\linewidth]{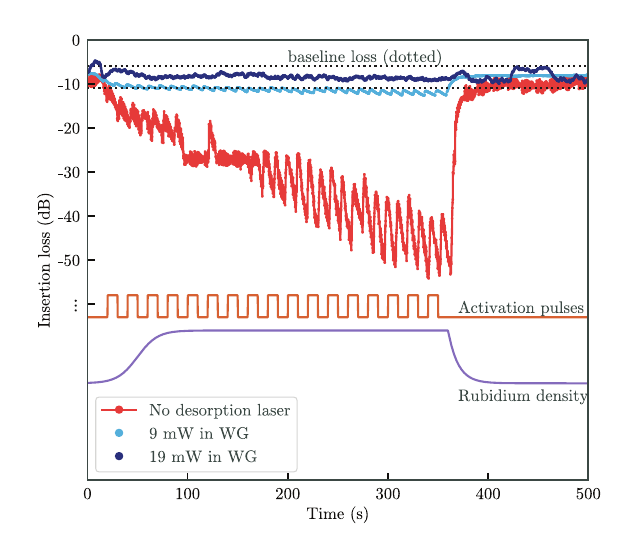}
    \caption{\textbf{Insertion loss of a PIC waveguide under low power pulsed Rb activation and in-situ desorption.} Total insertion loss at \SI{780}{nm}, away from Rb resonance, is monitored while a \SI{980}{nm} activation laser dispenses Rb vapor (configuration in Fig.\ref{fig:setup}(f)). When no activation laser is applied, the baseline insertion loss resides within the black dotted lines, corresponding to a baseline value of $\approx \SI{9(1)}{\decibel}$ ($\approx \SI{5}{\decibel}$ per facet), with the range of values due to variation in fiber-PIC coupling. The increase in Rb vapor density upon pill activation is directly responsible for the rise in insertion loss (red curve); when the activation laser is turned off, the released Rb is absorbed into the glass frame, and the Rb-induced loss reverses. A \SI{801}{nm} desorption laser is introduced to prevent Rb from sticking to the waveguide, with three settings compared: red curve (\SI{0}{mW} of desorption laser); light blue curve (\SI{9}{mW}); and dark blue curve (19~mW).  For full desorption laser power (dark blue curve), the waveguide insertion loss stays within the uncertainty of the baseline level, indicating essentially complete suppression of Rb-induced loss.}
    \label{fig:wgLoss}
\end{figure}

\begin{figure*}[ht]
    \centering
    \includegraphics[width=1\textwidth]{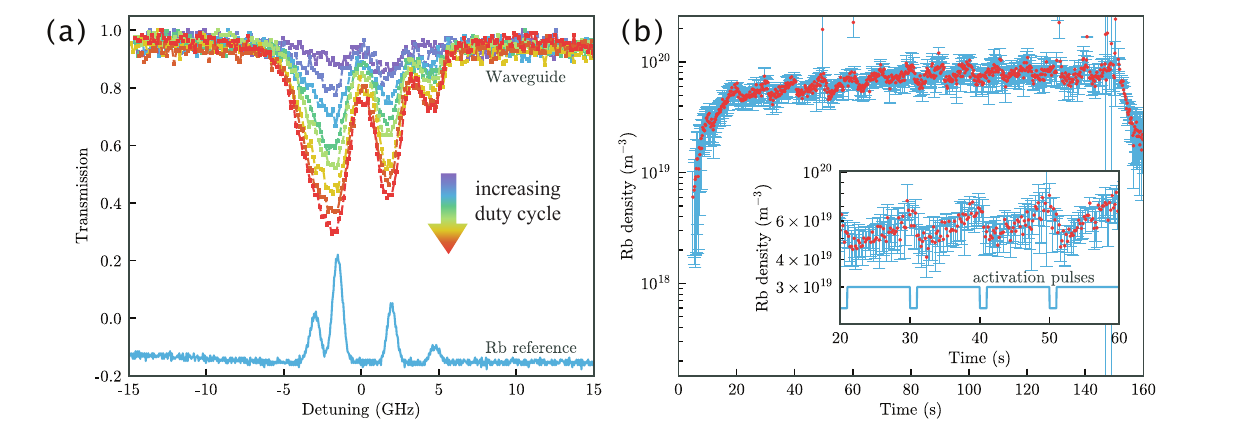}
    \caption{\textbf{Rubidium absorption in a 3 mm waveguide.}
    \textbf{(a)}\textit{Absorption spectra.}
Repetitive pulsing of a \SI{980}{nm} activation laser releases Rb from a pill while an \SI{801}{nm} desorption laser prevents Rb from coating the waveguide; the resulting transmission spectra of the $5S_{1/2} \rightarrow 5P_{3/2} $ manifold are recorded through a \SI{3}{mm} waveguide (Fig.~\ref{fig:setup}(f)). The multi-colored traces represent successively longer activation-laser duty cycles (see Fig. \ref{fig:RbCtrl}) and therefore progressively higher Rb densities. The traces exhibit characteristic broadening effects due to the short transit time of the Rb atoms through the evanescent mode of the waveguide. The lower blue trace is a reference spectrum recorded in a free-space cell with light split off from the same probe beam for comparison.
 \textbf{(b)} \textit{Temporal evolution of the Rb vapor density.} Absorption spectra are collected over a \SI{150}{s} interval while a \SI{980}{nm} activation laser (\SI{9}{s} ON / \SI{1}{s} OFF) periodically heats the Rb pill to generate Rb vapor. Densities extracted from each spectrum are plotted as red markers, with one standard deviation fit uncertainties shown by the blue band. The vapor density climbs toward a quasi-steady state during repeated heating pulses and decays once pill activation stops. \textbf{Inset:} expanded view of several pulses, highlighting the sawtooth rise during each \SI{9}{s} ON period and the fall during the \SI{1}{s} OFF period.
 }
    \label{fig:wgSpectra}
\end{figure*}
Next, we examine the use of low-power, pulsed activation of the Rb pill as described in the previous sections in experiments that probe the resulting Rb vapor using the evanescent field of a PIC waveguide. Our first consideration is the impact of the Rb vapor on waveguide insertion loss at \SI{780}{nm}, non-resonant with any Rb transition. We monitor the variation in insertion loss, as defined in section \ref{section:60micrometer}, as the Rb density in the vapor--PIC device increases with accumulated laser activation pulses, where the waveguide in use has a \SI{3}{mm} long air-clad section that is exposed to Rb. We activate the pill as discussed in section \ref{section:pulsed}, and compare behavior with and without the introduction of a counter-propagating 801 nm desorption laser.  

Before pill activation and Rb release into the physics chamber containing the photonics, the PIC waveguide has an average, baseline total loss of around $\approx \SI{9(1)}{\decibel}$ at \SI{780}{nm}, where the uncertainty is a one standard deviation value associated with fluctuations in the fiber-PIC coupling stability. We then monitor this insertion loss as we apply laser activation pulses to the Rb pill (without any desorption laser). The red sawtooth, dashed line in Fig.~\ref{fig:wgLoss} tracks the insertion loss of the waveguide; a depiction of the target activation pulses and transmission curve is placed below it. We see that the increase in Rb vapor is directly responsible for the increase in insertion loss as evident from the loss increasing and recovering (sawtooth pattern) as the activation pulses are turned on and off. This interpretation is further cemented by the total recovery of the transmission through the waveguide after the pill activation stops and Rb adsorbs onto the glass. The total insertion loss through the PIC waveguide increases by more than \SI{40}{\decibel} from the baseline value at maximum Rb density.

To combat this catastrophic increase in insertion loss, we use a desorption laser at \SI{801}{nm} from a diode laser with a maximum power output of $\approx \SI{60}{mW}$.  However, due to the baseline insertion loss, the total laser power in the waveguide is lower. Assuming symmetric loss for both input and output facets, we have approximately \SI{4}{\decibel} to \SI{5}{\decibel} loss for the input facet, so that the estimated desorption laser power in the waveguide with an input power of $\approx \SI{60}{mW}$ is $\approx \SI{19}{mW}$. At this desorption laser power level, we do not observe any sign of loss due to Rb (green data in Fig.~\ref{fig:wgLoss}). At a lower desorption power of $\approx \SI{9}{mW}$ in the waveguide, most of the loss due to Rb vapor introduction is avoided. The blue sawtooth dashed line in Fig.~\ref{fig:wgLoss} shows the resulting transmission characteristic, where the sawtooth pattern is reminiscent of the red curve (no desorption laser), but the amount of degradation is far less, with the maximum increase in loss within 1~dB to 2~dB of the baseline waveguide loss.

We note that the use of a desorption laser has been previously shown to be an effective deterrent against adverse Rb effects when introduced into the near-field of propagating modes in hollow core fibers \cite{sprague_efficient_2013,donvalkar_frequency_2014} and tapered nanofibers \cite{finkelstein_super-extended_2021}. Both material heating \cite{lai_transmission_2013} and light-induced atomic desorption (LIAD) \cite{bhagwat_low-light-level_2010} have been cited as mechanisms for alkali vapor desorption. In our case, it is not clear if it is one or a combination of both effects. A deeper understanding of the physical mechanism by which the desorption laser eliminates PIC waveguide degradation will be the subject of future work. 

With the desorption laser in place, the increase in propagation loss of the waveguide is minimized, which allows for stable waveguide-based spectroscopy of the Rb atoms. The strong confinement of probe light through the waveguide translates to much higher intensity of light experienced by the atoms compared to free-space, so it is much easier to saturate the atomic transitions. For this reason, we use a photoreceiver with sensitivity at the $<\SI{100}{fW}$ level to measure the 780 nm probe light, which is at a few nanowatts of power. This also means that, even with a counter-propagating configuration, the desorption laser reflection from the PIC facet of $\approx 0.5~\%$ of the incident power can easily overwhelm the photoreceiver. Hence, the output from the waveguide is passed through a 780 nm bandpass filter before it reaches the photoreceiver.  

   In Fig.~\ref{fig:wgSpectra}(a) we plot the waveguide transmission probing the  $5S_{1/2} \rightarrow 5P_{3/2} $ manifold through the 3 mm air-clad region while the Rb pill is pulsed to release Rb vapor and the desorption laser keeps Rb-induced deterioration to a minimum. Multicolored, stacked traces show the transmission spectra in which the Rb-induced absorption features have a contrast that increases with longer activation laser duty cycle and therefore progressively higher Rb densities. The traces were taken at $180~^\circ$C and constant activation laser power. An inverted free-space absorption spectrum from a reference vapor cell is shown at the bottom. The waveguide-mediated absorption features are similar to those previously observed in atomic vapor clad PIC waveguides~\cite{stern_evanescent_2013,ritter_atomic_2015}, where a combination of Doppler broadening (exacerbated by the waveguide's non-unity effective index) and transit time broadening (due to the short transit of the atoms through the evanescent field outside the waveguide) lead to spectral dips that are significantly broader than that of the Rb reference cell.

We can take many sequences of these absorption spectra at a particular duty cycle, temperature, and activation laser power as the Rb pill is pulsed to follow the temporal evolution of the vapor density, broadly similar to transmission plots shown in Figs.~\ref{fig:RbCtrl}(b) and \ref{fig:tempDepPulsed}.  We extract the Rb density for each spectrum in the sequence using ratios of the transmission dip depths of the Rb hyperfine transitions. Straightforward fitting of the spectra is complicated by the non-standard Doppler distribution of the atoms as discussed in section \ref{section:temp}. We discuss the density fitting methods further in the Appendix. 

The plot in Fig.~\ref{fig:wgSpectra}(b) follows the Rb density for \SI{160}{s} with the activation laser on from \SI{0}{s} to \SI{150}{s}. The duty cycle is set with $tON=\SI{9}{s}$ and $tOFF=\SI{1}{s}$. The extracted densities are plotted as red markers while the blue band shows one standard deviation fit uncertainties. As expected we see a fast increase in Rb density, a quick plateau, and a fast decay after the pulsing stops. The plot also exhibits the sawtooth pattern which appeared in the insertion loss through the waveguide in Fig.~\ref{fig:wgLoss}. An expanded view shown in the inset shows a clear relationship between each laser pulse and the Rb density; the sawtooth rises during each \SI{9}{s} $tON$ period and falls during the \SI{1}{s} $tOFF$ period. 

\section{Outlook}
In summary, we have demonstrated an approach that enables near-field vapor--PIC interactions in compact, manufacturable, and standalone devices that combine PICs with micromachined vapor cells and commercially available Rb pill sources. After observing that standard Rb pill activation led to irreversible and pronounced losses in air-clad PIC waveguides, we implemented a low-power activation method for controlled Rb release and used a desorption laser to prevent waveguide degradation. Using this approach, 3~mm long air-clad regions show no excess loss when exposed to Rb vapor, and waveguide-based spectroscopy of the Rb D2 transitions was successfully shown. In addition, we demonstrated control of Rb vapor density, enabling either transient bursts or quasi-steady states by adjusting the activation laser duty cycle and power, and we characterized the effect of temperature on the Rb density within the cell.

Going forward, we envision different applications that can leverage the transient Rb densities produced by our approach. For example, transient Rb density similar has previously been explored in photonic band-gap fibers~\cite{ghosh_low-light-level_2006} for electromagnetically induced transparency (EIT) and in thin vapor cells~\cite{christaller_transient_2022} to study dipole interactions. We further expect that this mode of operation will be particularly relevant for experiments such as single-photon sources in cavity QED, where regardless of the nature of the alkali source, atoms will transiently interact with the confined modes of photonic devices~\cite{zektzer_strong_2024,larsen_chip-scale_2025,austin_vapor-cavity-qed_2025}. In addition, being able to turn off the presence of Rb atoms entirely can be beneficial for experiment calibration and long-term device operation. On the other hand, transient Rb densities may not be suitable for applications that require or benefit from a stable and continuous vapor presence, such as frequency standards for laser locking. Future work will thus also focus on mitigating Rb adsorption onto glass through atomic layer deposition (ALD) of \ce{Al2O3}, which has been shown to significantly extend vapor cell lifetimes~\cite{woetzel_lifetime_2013}.

\medskip
\noindent\textbf{Funding}  The authors acknowledge partial funding support from the DARPA SAVaNT program through ARO contract W911NF2120106, the NIST-on-a-chip program, and a seedling grant from Northrop Grumman. 

\medskip
\noindent \textbf{Acknowledgements --} The authors thank Glenn Holland and Alessandro Restelli for experimental support and Alan Midgall for access to laboratory space.

\medskip
\noindent \textbf{Disclosures} The authors declare no conflicts of interest.

\medskip
\noindent \textbf{Data Availability} Data underlying the results presented in this paper are not publicly available at this time but may be obtained from the authors upon reasonable
request.

\clearpage

\appendix
\beginsupplement
\section{Determination of Rubidium Density}
The alkali vapor number density ($N$) is an essential parameter to determine optical depth in various experiments and is often required to model nonlinear interactions in quantum optics. Typical experiments use a vapor equilibrium model like the Clausius-Clapeyron (CC) law to determine $N$, assuming a known vapor temperature. Alternatively, vapor density can also be easily extracted by probing the transmission spectra and fitting the Doppler broadened lines, given that the length of the cell ($L$) is known. 

However, in our case, we have two main reasons why these common methods fail to produce reliable density extraction. 
First, due to the transient nature of the Rb vapor and compact geometry its velocity distribution has not been characterized sufficiently well to properly to fit the broadened lines accurately.
The second reason is associated with a fluctuating background signal, which though low in absolute power, still makes a significant impact. As outlined in the main text, we use a high power ($\approx\SI{10}{mW}$) desorption laser. Even in a counter-propagating configuration and with heavy filtering, facet reflection leaks into the signal, since a highly sensitive photoreceiver is used to measure the few nanowatts of probe light. One would assume the background from the desorption laser can be calibrated, but the back reflection is sensitive to the lensed fiber coupling which itself is prone to perturbations from heating effects of the activation laser. This leads to a fluctuating background signal that makes getting accurate transmission dips over a period of a measurement sequence lasting up to \SI{300}{s} very difficult.

In this Appendix we present a method to circumvent this problem. In short, we normalize the waveguide spectrum and compare the ratio of transmission dips at D2 lines of \ce{^{85}Rb} and \ce{^{87}Rb} against the CC-model to infer corresponding temperature ($T$). The steps, divided into two parts, are listed numerically in the following subsections.

\subsection{Processing the waveguide spectra}

\begin{figure}
    \centering
    \includegraphics[width=\linewidth]{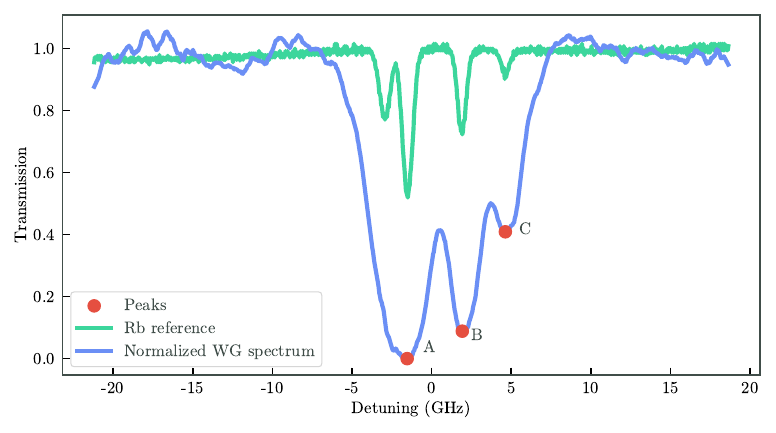}
    \caption{Plot showing a normalized Rb transmission spectra taken through the waveguide. The three red dots shows the visible dips associate with hyperfine transitions in \ce{85Rb} and \ce{87Rb}} labeled as A,B, and C. 
    \label{fig:peaks}
\end{figure}

\begin{enumerate}
    \item Normalize the spectra so that the transmission range is between 0 and 1 as shown in Fig. \ref{fig:peaks}. 
    \item Find the dips corresponding to $5S_{1/2},F=1 \rightarrow 5P_{3/2} $  and $5S_{1/2},F=2 \rightarrow 5P_{3/2} $  in \ce{^{85}Rb}, and $5S_{1/2},F=1 \rightarrow 5P_{3/2} $ transition in \ce{^{87}Rb}. We will call the minimum transmission values at these dips A,B, and C for convenience, labeled in Fig. \ref{fig:peaks}. 
    \item Find the values $R_1=1/(1-A+B)$ and $R_2=1/(1-A+C)$.
\end{enumerate}

\subsection{Using peak ratios to fit temperature}
\begin{figure}
    \centering
    \includegraphics[width=1\linewidth]{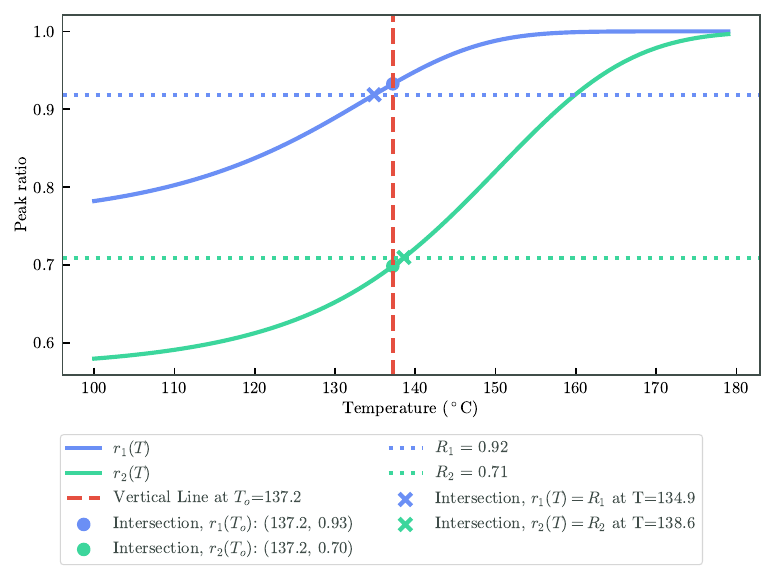}
    \caption{Two temperature dependent curves are generated using the CC-model to estimate equivalent temperature from recorded spectra.}
    \label{fig:fitting}
\end{figure}

\begin{enumerate}
    \item Generate quantities $r_1$ and $r_2$ analogous to values $R_1$ and $R_2$ respectively for different temperatures using the CC-model. These are the two blue and green curves in Fig. \ref{fig:fitting}.
    \item Given $R_1$ and $R_2$ values derived from the waveguide spectra like the one presented in Fig. \ref{fig:peaks}, find $T$ such that $r_1(T)=R_1$ and $r_2(T)=R_2$. This is visualized by drawing horizontal lines at $y=R_1$ and $y=R_2$ that intersect the curves $r_1$ and $r_2$ defined by the CC-model.
    \item Next find $T=T_o$ that minimizes the function $f(T)=\sqrt{(r_1(T)-R_1)^2 +(r_2(T)-R_2)^2}$. A vertical line constructed shows that it minimizes the total distance between both curves' actual values and the target values at $T_o$. 
    \item $T_o$ is then converted to a Rb vapor density using the CC-model. 
\end{enumerate}

We use this method to generate the plot shown in Fig. 8 of the paper. As the zoomed-in inset in the figure shows, the method is able to capture the effect of the activation laser on the relative density of Rb vapor in the device very accurately. However, the method has a few limitations. 
The first obvious limitation is that all three transitions need to exhibit a spectral dip whose amplitude exceeds the noise level.
The second, less apparent limitation comes from the narrow range over which the curves $r_1(T)$ and $r_2(T)$ are sensitive to the change in transmission depth. At both lower and higher temperatures the slope of the $r_1(T)$ and $r_2(T)$ becomes small such that small changes in transmission depth realize large temperature differences. These effects are reflected in the size of the error bars in Fig.~8 in the main text.


\begin{thebibliography}{10}
\newcommand{\enquote}[1]{``#1''}
\expandafter\ifx\csname url\endcsname\relax
  \def\url#1{\texttt{#1}}\fi
\expandafter\ifx\csname urlprefix\endcsname\relax\def\urlprefix{URL }\fi
\providecommand{\eprint}[2][]{\url{#2}}

\bibitem{kitching_chip-scale_2018}
J.~Kitching, \enquote{Chip-scale atomic devices,} Applied Physics Reviews \textbf{5}(3), 031,302 (2018). \urlprefix\url{https://doi.org/10.1063/1.5026238}.

\bibitem{spillane_observation_2008}
S.~M. Spillane, G.~S. Pati, K.~Salit, M.~Hall, P.~Kumar, R.~G. Beausoleil, and M.~S. Shahriar, \enquote{Observation of {Nonlinear} {Optical} {Interactions} of {Ultralow} {Levels} of {Light} in a {Tapered} {Optical} {Nanofiber} {Embedded} in a {Hot} {Rubidium} {Vapor},} Physical Review Letters \textbf{100}(23), 233,602 (2008). Publisher: American Physical Society, \urlprefix\url{https://link.aps.org/doi/10.1103/PhysRevLett.100.233602}.

\bibitem{salit_ultra-low_2011}
K.~Salit, M.~Salit, S.~Krishnamurthy, Y.~Wang, P.~Kumar, and M.~S. Shahriar, \enquote{Ultra-low power, {Zeno} effect based optical modulation in a degenerate {V}-system with a tapered nano fiber in atomic vapor,} Optics Express \textbf{19}(23), 22,874--22,881 (2011). Publisher: Optica Publishing Group, \urlprefix\url{https://opg.optica.org/oe/abstract.cfm?uri=oe-19-23-22874}.

\bibitem{jones_saturation_2014}
D.~E. Jones, J.~D. Franson, and T.~B. Pittman, \enquote{Saturation of atomic transitions using subwavelength diameter tapered optical fibers in rubidium vapor,} Journal of the Optical Society of America B \textbf{31}(8), 1997 (2014). \urlprefix\url{https://opg.optica.org/josab/abstract.cfm?uri=josab-31-8-1997}.

\bibitem{jones_ladder-type_2015}
D.~E. Jones, J.~D. Franson, and T.~B. Pittman, \enquote{Ladder-type electromagnetically induced transparency using nanofiber-guided light in a warm atomic vapor,} Physical Review A \textbf{92}(4), 043,806 (2015). Publisher: American Physical Society, \urlprefix\url{https://link.aps.org/doi/10.1103/PhysRevA.92.043806}.

\bibitem{song_absorption_2019}
Z.~Song, X.~Yue, Y.~Luo, H.~Li, and Y.~Zhao, \enquote{Absorption saturation measurement using the tapered optical nanofiber in a hot cesium vapor,} Chinese Optics Letters \textbf{17}(3), 031,901 (2019). Publisher: Chinese Optical Society, \urlprefix\url{https://opg.optica.org/col/abstract.cfm?uri=col-17-3-031901}.

\bibitem{lamsal_transmission_2019}
H.~P. Lamsal, J.~D. Franson, and T.~B. Pittman, \enquote{Transmission characteristics of optical nanofibers in metastable xenon,} Applied Optics \textbf{58}(24), 6470--6473 (2019). Publisher: Optica Publishing Group, \urlprefix\url{https://opg.optica.org/ao/abstract.cfm?uri=ao-58-24-6470}.

\bibitem{finkelstein_super-extended_2021}
R.~Finkelstein, G.~Winer, D.~Z. Koplovich, O.~Arenfrid, T.~Hoinkes, G.~Guendelman, M.~Netser, E.~Poem, A.~Rauschenbeutel, B.~Dayan, and O.~Firstenberg, \enquote{Super-extended nanofiber-guided field for coherent interaction with hot atoms,} Optica \textbf{8}(2), 208--215 (2021). Publisher: Optica Publishing Group, \urlprefix\url{https://opg.optica.org/optica/abstract.cfm?uri=optica-8-2-208}.

\bibitem{ghosh_low-light-level_2006}
S.~Ghosh, A.~R. Bhagwat, C.~K. Renshaw, S.~Goh, A.~L. Gaeta, and B.~J. Kirby, \enquote{Low-{Light}-{Level} {Optical} {Interactions} with {Rubidium} {Vapor} in a {Photonic} {Band}-{Gap} {Fiber},} Physical Review Letters \textbf{97}(2), 023,603 (2006). \urlprefix\url{https://link.aps.org/doi/10.1103/PhysRevLett.97.023603}.

\bibitem{londero_ultralow-power_2009}
P.~Londero, V.~Venkataraman, A.~R. Bhagwat, A.~D. Slepkov, and A.~L. Gaeta, \enquote{Ultralow-{Power} {Four}-{Wave} {Mixing} with {Rb} in a {Hollow}-{Core} {Photonic} {Band}-{Gap} {Fiber},} Physical Review Letters \textbf{103}(4), 043,602 (2009). Publisher: American Physical Society, \urlprefix\url{https://link.aps.org/doi/10.1103/PhysRevLett.103.043602}.

\bibitem{slepkov_spectroscopy_2010}
A.~D. Slepkov, A.~R. Bhagwat, V.~Venkataraman, P.~Londero, and A.~L. Gaeta, \enquote{Spectroscopy of {Rb} atoms in hollow-core fibers,} Physical Review A \textbf{81}(5), 053,825 (2010). Publisher: American Physical Society, \urlprefix\url{https://link.aps.org/doi/10.1103/PhysRevA.81.053825}.

\bibitem{venkataraman_few-photon_2011}
V.~Venkataraman, K.~Saha, P.~Londero, and A.~L. Gaeta, \enquote{Few-{Photon} {All}-{Optical} {Modulation} in a {Photonic} {Band}-{Gap} {Fiber},} Physical Review Letters \textbf{107}(19), 193,902 (2011). Publisher: American Physical Society, \urlprefix\url{https://link.aps.org/doi/10.1103/PhysRevLett.107.193902}.

\bibitem{sprague_efficient_2013}
M.~R. Sprague, D.~G. England, A.~Abdolvand, J.~Nunn, X.-M. Jin, W.~S. Kolthammer, M.~Barbieri, B.~Rigal, P.~S. Michelberger, T.~F.~M. Champion, P.~S.~J. Russell, and I.~A. Walmsley, \enquote{Efficient optical pumping and high optical depth in a hollow-core photonic-crystal fibre for a broadband quantum memory,} New Journal of Physics \textbf{15}(5), 055,013 (2013). Publisher: IOP Publishing, \urlprefix\url{https://dx.doi.org/10.1088/1367-2630/15/5/055013}.

\bibitem{perrella_high-resolution_2013}
C.~Perrella, P.~S. Light, J.~D. Anstie, T.~M. Stace, F.~Benabid, and A.~N. Luiten, \enquote{High-resolution two-photon spectroscopy of rubidium within a confined geometry,} Physical Review A \textbf{87}(1), 013,818 (2013). Publisher: American Physical Society, \urlprefix\url{https://link.aps.org/doi/10.1103/PhysRevA.87.013818}.

\bibitem{donvalkar_frequency_2014}
P.~S. Donvalkar, V.~Venkataraman, S.~Clemmen, K.~Saha, and A.~L. Gaeta, \enquote{Frequency translation via four-wave mixing {Bragg} scattering in {Rb} filled photonic bandgap fibers,} Optics Letters \textbf{39}(6), 1557--1560 (2014). Publisher: Optica Publishing Group, \urlprefix\url{https://opg.optica.org/ol/abstract.cfm?uri=ol-39-6-1557}.

\bibitem{perrella_engineering_2018}
C.~Perrella, P.~Light, S.~A. Vahid, F.~Benabid, and A.~Luiten, \enquote{Engineering {Photon}-{Photon} {Interactions} within {Rubidium}-{Filled} {Waveguides},} Physical Review Applied \textbf{9}(4), 044,001 (2018). Publisher: American Physical Society, \urlprefix\url{https://link.aps.org/doi/10.1103/PhysRevApplied.9.044001}.

\bibitem{schmidt_atomic_2010}
H.~Schmidt and A.~Hawkins, \enquote{Atomic spectroscopy and quantum optics in hollow-core waveguides,} Laser \& Photonics Reviews \textbf{4}(6), 720--737 (2010). \_eprint: https://onlinelibrary.wiley.com/doi/pdf/10.1002/lpor.200900040, \urlprefix\url{https://onlinelibrary.wiley.com/doi/abs/10.1002/lpor.200900040}.

\bibitem{yang_atomic_2007}
W.~Yang, D.~B. Conkey, B.~Wu, D.~Yin, A.~R. Hawkins, and H.~Schmidt, \enquote{Atomic spectroscopy on a chip,} Nature Photonics \textbf{1}(6), 331--335 (2007). Publisher: Nature Publishing Group, \urlprefix\url{https://www.nature.com/articles/nphoton.2007.74}.

\bibitem{wu_slow_2010}
B.~Wu, J.~F. Hulbert, E.~J. Lunt, K.~Hurd, A.~R. Hawkins, and H.~Schmidt, \enquote{Slow light on a chip via atomic quantum state control,} Nature Photonics \textbf{4}(11), 776--779 (2010). Publisher: Nature Publishing Group, \urlprefix\url{https://www.nature.com/articles/nphoton.2010.211}.

\bibitem{stern_nanoscale_2013}
L.~Stern, B.~Desiatov, I.~Goykhman, and U.~Levy, \enquote{Nanoscale light–matter interactions in atomic cladding waveguides,} Nature Communications \textbf{4}(1), 1548 (2013). Publisher: Nature Publishing Group, \urlprefix\url{https://www.nature.com/articles/ncomms2554}.

\bibitem{ritter_atomic_2015}
R.~Ritter, N.~Gruhler, W.~Pernice, H.~Kübler, T.~Pfau, and R.~Löw, \enquote{Atomic vapor spectroscopy in integrated photonic structures,} Applied Physics Letters \textbf{107}(4), 041,101 (2015). \urlprefix\url{https://doi.org/10.1063/1.4927172}.

\bibitem{stern_strong_2017}
L.~Stern, B.~Desiatov, N.~Mazurski, and U.~Levy, \enquote{Strong coupling and high-contrast all-optical modulation in atomic cladding waveguides,} Nature Communications \textbf{8}(1), 14,461 (2017). Number: 1 Publisher: Nature Publishing Group, \urlprefix\url{https://www.nature.com/articles/ncomms14461}.

\bibitem{ritter_coupling_2018}
R.~Ritter, N.~Gruhler, H.~Dobbertin, H.~Kübler, S.~Scheel, W.~Pernice, T.~Pfau, and R.~Löw, \enquote{Coupling {Thermal} {Atomic} {Vapor} to {Slot} {Waveguides},} Physical Review X \textbf{8}(2), 021,032 (2018). Publisher: American Physical Society, \urlprefix\url{https://link.aps.org/doi/10.1103/PhysRevX.8.021032}.

\bibitem{zektzer_nanoscale_2021}
R.~Zektzer, N.~Mazurski, Y.~Barash, and U.~Levy, \enquote{Nanoscale atomic suspended waveguides for improved vapour coherence times and optical frequency referencing,} Nature Photonics \textbf{15}(10), 772--779 (2021). Publisher: Nature Publishing Group, \urlprefix\url{https://www.nature.com/articles/s41566-021-00853-4}.

\bibitem{zektzer_strong_2024}
R.~Zektzer, X.~Lu, K.~T. Hoang, R.~Shrestha, S.~Austin, F.~Zhou, A.~Chanana, G.~Holland, D.~Westly, P.~Lett, A.~V. Gorshkov, and K.~Srinivasan, \enquote{Strong interactions between integrated microresonators and alkali atomic vapors: towards single-atom, single-photon operation,} Optica \textbf{11}(10), 1376--1384 (2024). Publisher: Optica Publishing Group, \urlprefix\url{https://opg.optica.org/optica/abstract.cfm?uri=optica-11-10-1376}.

\bibitem{davidson-marquis_coherent_2021}
F.~Davidson-Marquis, J.~Gargiulo, E.~Gómez-López, B.~Jang, T.~Kroh, C.~Müller, M.~Ziegler, S.~A. Maier, H.~Kübler, M.~A. Schmidt, and O.~Benson, \enquote{Coherent interaction of atoms with a beam of light confined in a light cage,} Light: Science \& Applications \textbf{10}(1), 114 (2021). Publisher: Nature Publishing Group, \urlprefix\url{https://www.nature.com/articles/s41377-021-00556-z}.

\bibitem{poon_silicon_2024}
J.~K.~S. Poon, A.~Govdeli, A.~Sharma, X.~Mu, F.-D. Chen, T.~Xue, and T.~Liu, \enquote{Silicon photonics for the visible and near-infrared spectrum,} Advances in Optics and Photonics \textbf{16}(1), 1 (2024). \urlprefix\url{https://opg.optica.org/abstract.cfm?URI=aop-16-1-1}.

\bibitem{perez_high-performance_2023}
E.~F. Perez, G.~Moille, X.~Lu, J.~Stone, F.~Zhou, and K.~Srinivasan, \enquote{High-performance {Kerr} microresonator optical parametric oscillator on a silicon chip,} Nature Communications \textbf{14}(1), 242 (2023). Publisher: Nature Publishing Group, \urlprefix\url{https://www.nature.com/articles/s41467-022-35746-9}.

\bibitem{moille_integrated_2022}
G.~Moille, D.~Westly, E.~F. Perez, M.~Metzler, G.~Simelgor, and K.~Srinivasan, \enquote{Integrated buried heaters for efficient spectral control of air-clad microresonator frequency combs,} APL Photonics \textbf{7}(12), 126,104 (2022). \urlprefix\url{https://doi.org/10.1063/5.0127466}.

\bibitem{zhou_prospects_2023}
Z.~Zhou, X.~Ou, Y.~Fang, E.~Alkhazraji, R.~Xu, Y.~Wan, and J.~E. Bowers, \enquote{Prospects and applications of on-chip lasers,} eLight \textbf{3}(1), 1 (2023). \urlprefix\url{https://doi.org/10.1186/s43593-022-00027-x}.

\bibitem{isichenko_multi-laser_2025}
A.~Isichenko, A.~S. Hunter, N.~Chauhan, J.~R. Dickson, T.~N. Nunley, J.~R. Bingaman, D.~A.~S. Heim, M.~W. Harrington, K.~Liu, P.~D. Kunz, and D.~J. Blumenthal, \enquote{Multi-laser stabilization with an atomic-disciplined photonic integrated resonator,}  (2025). ArXiv:2509.09124 [physics], \urlprefix\url{http://arxiv.org/abs/2509.09124}.

\bibitem{riley_evanescent_2025}
P.~Riley, K.~T. Hoang, R.~Shrestha, R.~Zektzer, D.~Westly, K.~Srinivasan, and M.~Hummon, \enquote{Evanescent {Light}-{Matter} {Interaction} in an {Integrated} {MEMS} – {Nanophotonic} {Vapor} {Cell},} in \emph{{CLEO} 2025 (2025), paper {SS117}\_6}, p. SS117\_6 (Optica Publishing Group, 2025). \urlprefix\url{https://opg.optica.org/abstract.cfm?uri=CLEO_SI-2025-SS117_6}.

\bibitem{grosman_wafer-scale_2025}
A.~Grosman, R.~Zektzer, N.~Mazurski, L.~Stern, and U.~Levy, \enquote{Wafer-scale integration of photonic integrated circuits and atomic vapor cells,} Nanophotonics  (2025). Publisher: De Gruyter. DOI: 10.1515/nanoph-2025-0500, \urlprefix\url{https://www.degruyterbrill.com/document/doi/10.1515/nanoph-2025-0500/html}.

\bibitem{loh_microresonator_2016}
W.~Loh, M.~T. Hummon, H.~F. Leopardi, T.~M. Fortier, F.~Quinlan, J.~Kitching, S.~B. Papp, and S.~A. Diddams, \enquote{Microresonator {Brillouin} laser stabilization using a microfabricated rubidium cell,} Optics Express \textbf{24}(13), 14,513--14,524 (2016). Publisher: Optica Publishing Group, \urlprefix\url{https://opg.optica.org/oe/abstract.cfm?uri=oe-24-13-14513}.

\bibitem{sebbag_chip-scale_2021}
Y.~Sebbag, A.~Naiman, E.~Talker, Y.~Barash, and U.~Levy, \enquote{Chip-{Scale} {Integration} of {Nanophotonic}-{Atomic} {Magnetic} {Sensors},} ACS Photonics \textbf{8}(1), 142--146 (2021). Publisher: American Chemical Society, \urlprefix\url{https://doi.org/10.1021/acsphotonics.0c01473}.

\bibitem{knapkiewicz_dynamically_2019}
P.~Knapkiewicz and T.~Grzebyk, \enquote{Dynamically stabilized high vacuum inside rubidium vapor {MEMS} cell for cold atom spectroscopy,} in \emph{2019 19th {International} {Conference} on {Micro} and {Nanotechnology} for {Power} {Generation} and {Energy} {Conversion} {Applications} ({PowerMEMS})}, pp. 1--6 (2019). \urlprefix\url{https://ieeexplore.ieee.org/document/9080303/}.

\bibitem{han_microfabricated_2018}
R.~Han, Z.~You, F.~Zhang, H.~Xue, and Y.~Ruan, \enquote{Microfabricated {Vapor} {Cells} with {Reflective} {Sidewalls} for {Chip} {Scale} {Atomic} {Sensors},} Micromachines \textbf{9}(4), 175 (2018). Publisher: Multidisciplinary Digital Publishing Institute, \urlprefix\url{https://www.mdpi.com/2072-666X/9/4/175}.

\bibitem{hummon_photonic_2018}
M.~T. Hummon, S.~Kang, D.~G. Bopp, Q.~Li, D.~A. Westly, S.~Kim, C.~Fredrick, S.~A. Diddams, K.~Srinivasan, V.~Aksyuk, and J.~E. Kitching, \enquote{Photonic chip for laser stabilization to an atomic vapor with 10$^{\textrm{−11}}$ instability,} Optica \textbf{5}(4), 443--449 (2018). Publisher: Optica Publishing Group, \urlprefix\url{https://opg.optica.org/optica/abstract.cfm?uri=optica-5-4-443}.

\bibitem{stern_chip-scale_2019}
L.~Stern, D.~G. Bopp, S.~A. Schima, V.~N. Maurice, and J.~E. Kitching, \enquote{Chip-scale atomic diffractive optical elements,} Nature Communications \textbf{10}(1), 3156 (2019). Publisher: Nature Publishing Group, \urlprefix\url{https://www.nature.com/articles/s41467-019-11145-5}.

\bibitem{lucivero_laser-written_2022}
V.~G. Lucivero, A.~Zanoni, G.~Corrielli, R.~Osellame, and M.~W. Mitchell, \enquote{Laser-written vapor cells for chip-scale atomic sensing and spectroscopy,} Optics Express \textbf{30}(15), 27,149--27,163 (2022). Publisher: Optica Publishing Group, \urlprefix\url{https://opg.optica.org/oe/abstract.cfm?uri=oe-30-15-27149}.

\bibitem{januszewicz_chip-scale_2025}
J.~Januszewicz, A.~P. McWilliam, S.~Dyer, J.~P. McGilligan, P.~F. Griffin, E.~Riis, E.~Di~Gaetano, M.~Sorel, D.~J. Paul, and K.~Gallacher, \enquote{Chip-scale atomic spectrometer with silicon nitride optical phased array,} APL Photonics \textbf{10}(7), 076,104 (2025). \urlprefix\url{https://doi.org/10.1063/5.0273108}.

\bibitem{alaeian_cavity_2020}
H.~Alaeian, R.~Ritter, M.~Basic, R.~Löw, and T.~Pfau, \enquote{Cavity {QED} based on room temperature atoms interacting with a photonic crystal cavity: a feasibility study,} Applied Physics B \textbf{126}(2), 25 (2020). \urlprefix\url{https://doi.org/10.1007/s00340-019-7367-9}.

\bibitem{austin_vapor-cavity-qed_2025}
S.~Austin, D.~Devulapalli, K.~Hoang, F.~Zhou, K.~Srinivasan, and A.~V. Gorshkov, \enquote{A vapor-cavity-{QED} system for quantum computation and communication,}  (2025). ArXiv:2509.19432 [quant-ph], \urlprefix\url{http://arxiv.org/abs/2509.19432}.

\bibitem{larsen_chip-scale_2025}
B.~J. Larsen, H.~Hensley, G.~D. Martinez, A.~Staron, W.~R. McGehee, J.~Kitching, and J.~K. Thompson, \enquote{A chip-scale atomic beam source for non-classical light,}  (2025). ArXiv:2506.00199 [physics], \urlprefix\url{http://arxiv.org/abs/2506.00199}.

\bibitem{riley_manuscript_2025}
P.~Riley, R.~Shrestha, K.~Hoang, K.~Srinivasan, and M.~Hummon, \enquote{Manuscript in preparation,} unpublished  (2025).

\bibitem{stern_evanescent_2013}
L.~Stern, B.~Desiatov, I.~Goykhman, and U.~Levy, \enquote{Evanescent light-matter {Interactions} in {Atomic} {Cladding} {Wave} {Guides},} Nature Communications \textbf{4}(1), 1548 (2013). ArXiv:1204.0393 [physics], \urlprefix\url{http://arxiv.org/abs/1204.0393}.

\bibitem{stern_enhanced_2016}
L.~Stern, R.~Zektzer, N.~Mazurski, and U.~Levy, \enquote{Enhanced light-vapor interactions and all optical switching in a chip scale micro-ring resonator coupled with atomic vapor,} Laser \& Photonics Reviews \textbf{10}(6), 1016--1022 (2016). \_eprint: https://onlinelibrary.wiley.com/doi/pdf/10.1002/lpor.201600176, \urlprefix\url{https://onlinelibrary.wiley.com/doi/abs/10.1002/lpor.201600176}.

\bibitem{skljarow_purcell-enhanced_2022}
A.~Skljarow, H.~Kübler, C.~S. Adams, T.~Pfau, R.~Löw, and H.~Alaeian, \enquote{Purcell-enhanced dipolar interactions in nanostructures,} Physical Review Research \textbf{4}(2), 023,073 (2022). Publisher: American Physical Society, \urlprefix\url{https://link.aps.org/doi/10.1103/PhysRevResearch.4.023073}.

\bibitem{mcbride_demonstration_2025}
S.~E. McBride, C.~M. Gentry, C.~Holland, C.~Bellew, K.~R. Moore, and A.~Braun, \enquote{Demonstration of atom interrogation using photonic integrated circuits anodically bonded to ultra-high vacuum envelopes for epoxy-free scalable quantum sensors,} Optica Quantum \textbf{3}(1), 22--27 (2025). Publisher: Optica Publishing Group, \urlprefix\url{https://opg.optica.org/opticaq/abstract.cfm?uri=opticaq-3-1-22}.

\bibitem{douahi_vapour_2007}
A.~Douahi, L.~Nieradko, J.~Beugnot, J.~Dziuban, H.~Maillote, S.~Guérandel, M.~Moraja, C.~Gorecki, and V.~Giordano, \enquote{Vapour microcell for chip scale atomic frequency standard,} Electronics Letters \textbf{43}(5), 279--280 (2007). Publisher: The Institution of Engineering and Technology, \urlprefix\url{https://digital-library.theiet.org/doi/10.1049/el%3A20070147}.

\bibitem{giridhar_mems_2022}
M.~S. Giridhar, M.~M. Nandakishor, A.~Dahake, P.~Tiwari, A.~Jambhalikar, J.~John, and S.~P. Karanth, \enquote{{MEMS} {Rubidium} atomic vapor cell for miniature atomic resonance device applications,} ISSS Journal of Micro and Smart Systems \textbf{11}(2), 427--434 (2022). \urlprefix\url{https://doi.org/10.1007/s41683-022-00098-5}.

\bibitem{jia_microfabricated_2022}
S.~Jia, Z.~Jiang, B.~Jiao, X.~Liu, Y.~Pan, Z.~Song, and J.~Qu, \enquote{The {Microfabricated} {Alkali} {Vapor} {Cell} with {High} {Hermeticity} for {Chip}-{Scale} {Atomic} {Clock},} Applied Sciences \textbf{12}(1), 436 (2022). Number: 1 Publisher: Multidisciplinary Digital Publishing Institute, \urlprefix\url{https://www.mdpi.com/2076-3417/12/1/436}.

\bibitem{ma_modification_2009}
J.~Ma, A.~Kishinevski, Y.-Y. Jau, C.~Reuter, and W.~Happer, \enquote{Modification of glass cell walls by rubidium vapor,} Physical Review A \textbf{79}(4), 042,905 (2009). Publisher: American Physical Society, \urlprefix\url{https://link.aps.org/doi/10.1103/PhysRevA.79.042905}.

\bibitem{burnham_use_2017}
A.~K. Burnham, \enquote{Use and misuse of logistic equations for modeling chemical kinetics,} Journal of Thermal Analysis and Calorimetry \textbf{127}(1), 1107--1116 (2017). Company: Springer Distributor: Springer Institution: Springer Label: Springer Publisher: Springer Netherlands, \urlprefix\url{https://link.springer.com/article/10.1007/s10973-015-4879-3}.

\bibitem{gianni_di_domenico_vapor_2011}
{Gianni Di Domenico} and {Antoine Weis}, \enquote{Vapor {Pressure} and {Density} of {Alkali} {Metals},}  (2011). \urlprefix\url{http://demonstrations.wolfram.com/VaporPressureAndDensityOfAlkaliMetals/}.

\bibitem{lai_transmission_2013}
M.~Lai, J.~D. Franson, and T.~B. Pittman, \enquote{Transmission degradation and preservation for tapered optical fibers in rubidium vapor,} Applied Optics \textbf{52}(12), 2595--2601 (2013). Publisher: Optica Publishing Group, \urlprefix\url{https://opg.optica.org/ao/abstract.cfm?uri=ao-52-12-2595}.

\bibitem{bhagwat_low-light-level_2010}
A.~R. Bhagwat, \enquote{Low-{Light}-{Level} {Optical} {Interactions} with {Rubidium} {Vapor} in a {Photonic} {Band}-{Gap} {Fiber},} Ph.D. thesis, Cornell University (2010).

\bibitem{christaller_transient_2022}
F.~Christaller, M.~Mäusezahl, F.~Moumtsilis, A.~Belz, H.~Kübler, H.~Alaeian, C.~S. Adams, R.~Löw, and T.~Pfau, \enquote{Transient {Density}-{Induced} {Dipolar} {Interactions} in a {Thin} {Vapor} {Cell},} Physical Review Letters \textbf{128}(17), 173,401 (2022). \urlprefix\url{https://link.aps.org/doi/10.1103/PhysRevLett.128.173401}.

\bibitem{woetzel_lifetime_2013}
S.~Woetzel, F.~Talkenberg, T.~Scholtes, R.~IJsselsteijn, V.~Schultze, and H.-G. Meyer, \enquote{Lifetime improvement of micro-fabricated alkali vapor cells by atomic layer deposited wall coatings,} Surface and Coatings Technology \textbf{221}, 158--162 (2013). \urlprefix\url{https://linkinghub.elsevier.com/retrieve/pii/S0257897213001217}.

\end{thebibliography}
\end{document}